# Infinite Simple 3D Cubic Lattice of Identical Resistors with Two Missing Bonds

R. S. Hijjawi[1],   J. H. Asad [2] , A. J. Sakaji[3] , and  M. Al-sabayleh[1],
and J. M. Khalifeh [4] .

[1] Dep. of Physics – Mutah University, Karak- Jordan.

[2] Tabuk University,  Teachers College, Department of Sciences(Physics),  P.O.Box.1144,
Saudi Arabia.
  e-mail: jhasad1@yahoo.com.

[3] Dep. of Basic Sciences- Ajman University, Ajman- UAE.

[4] Department of Physics, University of Jordan, Amman-11942, Jordan.

**Abstract**

An infinite regular three-dimensional network is composed of identical resistors each of resistance $R$ joining adjacent nodes. What is the equivalent resistance between the lattice site $\vec{r}_i$ and the lattice $\vec{r}_j$ site, when two bonds are removed from the perfect network? Three cases are considered here, and some numerical values are calculated. Finally, the asymptotic behavior of the equivalent resistance is studied for large distances between the two sites.

## I- Introduction

A classic problem in electric circuit theory studied by numerous authors over many years is the computation of the resistance between two nodes in a resistor network (for a list of relevant references up to 2000 see, e.g., Refs.[1,2]). Besides being a central problem in electric circuit theory, the computation of resistances is also relevant to a wide range of problems ranging from random walks[3,4], the theory of harmonic functions[5], first-passage processes[6], to lattice Green's functions (LGF)[1,2,7]. The connection with these problems originates from the fact that electrical potentials on a grid are governed by the same difference equations as those occurring in the other problems. For this reason, the resistance problem is often studied from the point of view of solving the difference equations, which is most conveniently carried out for infinite networks. In the case of LGF approach which is presented by Cserti[1,7], efforts have been focused mainly on infinite lattices.

From the year 2004, the problem of calculating the equivalent resistance between two nodes in a resistor networks arises again in many papers. For example see the below efforts:

a- Asad[2] and Asad et. al[8,9]. studied the problem of calculating the equivalent resistance between any two lattice sites, using Cserti's method for both the square and Simple Cubic Lattices (SC). In there work two cases (i.e. perfect and perturbed (i.e. one bond is removed)) are investigated numerically and analytically, in addition to an experimental investigation. There was a good agreement between the mathematical and the experimental results especially for the bulk values. Also, there was a good agreement between their calculated values and those calculated by other previous authors[1,7,10–12].

b- Asad et. al[13]. extended the infinite square perturbed network where two bonds are removed. Numerical and analytical results are obtained.

Osterberg and Inan[14] studied the impedance between adjacent lattice sites for infinite D- dimensional resistive lattices where they show how one can find the total effective resistance between two adjacent sites of any D- dimensional resistive lattice.

Little attention has been paid to finite networks, even though the latter are those occurring in real life. Wu[15] took up this problem and present a general formulation for computing two-point resistances in finite networks. Particularly, he showed how to obtain the resistance between two arbitrary nodes in a resistor network in terms of the eigenvalues and eigenfunctions of the Laplacian matrix associated with the network. Explicit formulae for two-point resistances were deduced for regular lattices in one, two and three dimensions under various boundary conditions including that of a Mobius strip and a Klein bottle.

The LGF for cubic lattices has been investigated by many authors[16-27], and the so-called recurrence formulae which are often used to calculate the LGF of the SC at different sites are presented[19,20]. The values of the LGF for the SC lattice have been recently exactly evaluated[24], where these values are expressed in terms of the known value of the LGF at the origin. The LGF defined in our work is related to the Green's Function (GF) of the tight-binding Hamiltonian (TBH)[28].

In this work, we studied the perturbed infinite SC lattice when two bonds from the infinite SC network are removed using the LGF method presented by Cserti[1,7]. Numerical results are obtained and a comparison with those of the perfect infinite SC lattice and with those of the perturbed (i.e. due to removing one bond) is carried out. Also, the asymptotic behavior is investigated for large separation between the sites in the perturbed SC lattice.

## II- Perfect Case

Consider a perfect SC network consisting of identical resistors each with resistance R, and assume that all the lattice points to be specified by the position vector $\vec{r}$ given in the form

$$\vec{r} = l_1 \vec{a}_1 + l_2 \vec{a}_2 + ... + l_d \vec{a}_d . \tag{1}$$

where $l_1, l_2, ..., l_d$ are integers (positive, negative or zero),
and $\vec{a}_1, \vec{a}_2, ..., \vec{a}_d$ are independent primitive translation vectors.

If all the primitive translation vectors have the same magnitude, i.e., $|\vec{a}_1| = |\vec{a}_2| = ... = |\vec{a}_d| = a$, then the lattice is called hypercubic lattice. Here *a* is the lattice constant of the d-dimensional hypercube.

The equivalent resistance between the origin (0,0,0) and any other lattice site *(l,m,n)* has been expressed as[7]:

$$R_o(l,m,n) = R[G_o(3;0,0,0) - G_o(3;l,m,n)] . \tag{2}$$

where in a previous studied[1,7] it has been assumed that a current $(+I)$ enters at the origin and a current $(-I)$ exits at a lattice point $\vec{r}$, and zero otherwise. Thus:

$$I(\vec{r}') = \begin{cases} + I, & \vec{r}' = 0 \\ - I, & \vec{r}' = \vec{r} \\ 0, & \text{otherwise.} \end{cases} \qquad (3)$$

Also they[1,7] took the potential at the lattice point $\vec{r}'$ to be $V(\vec{r}')$.

It has been shown by Glasser et. al[29]. that the LGF of the SC lattice at any site $(l,m,n)$ can be expressed rationally in terms of $G_o(3;0,0,0)$ (i.e. LGF at the origin as):

$$G_o(3;l,m,n) = r_1 g_o + \frac{r_2}{\pi^2 g_o} + r_3. \qquad (4)$$

Substituting Eq. (4) into Eq. (2), one gets:

$$\frac{R_o(l,m,n)}{R} = \rho_1 g_o + \frac{\rho_2}{\pi^2 g_o} + \rho_3. \qquad (5)$$

where $g_o$ is the LGF of the infinite SC lattice at the origin. (i.e. $g_o = G_o(3;0,0,0) = 0.505462$.

$\rho_1, \rho_2$ and $\rho_3$ are rational numbers related to $r_1, r_2$ and $r_3$ (i.e. Duffin and Shelly's parameter[30]) as:

$$\rho_1 = 1 - r_1 = 1 - \lambda_1 - \frac{15}{12}\lambda_2;$$

$$\rho_2 = -r_2 = \frac{1}{2}\lambda_2;$$

and

$$\rho_3 = r_3 = \frac{1}{3}\lambda_3. \qquad (6)$$

Various values of $\rho_1, \rho_2$ and $\rho_3$ are presented in Ref[8]. and other values can be calculated using the following recurrence relation:

$$G_0(E;l+1,m,n) + G_0(E;l-1,m,n) + G_0(E;l,m+1,n) + G_0(E;l,m-1,n) +$$

$$G_0(E;l,m,n+1) + G_0(E;l,m,n-1) = -2\delta_{l0}\delta_{m0}\delta_{n0} + 2EG_0(E;l,m,n)$$

$$(7)$$

where $E = 3$, is the energy of the infinite SC lattice at the band.

In some cases one may use the above Eq. (7) two or three times to calculate different values of $\rho_1, \rho_2$ and $\rho_3$. Some calculated values for $R_o(l,m,n)$ are quoted in Table 1 below for comparison reason.

Finally, as the separation between the origin and the lattice site *(l,m,n)* goes to infinity then Eq. (2) becomes [1,2]:

$$\frac{R_o(l,m,n)}{R} \to g_o = 0.505462. \tag{8}$$

The resistance between the origin and any lattice site *(l,m,n)* in a perfect SC lattice goes to a finite value for large separation between the two sites.

### III- Perturbed SC Network (Two Resistors are Missing)

In this section, consider again the perfect infinite SC network specified in section II. Our aim here is to find the equivalent resistance between the site $i = (i_x, i_y, i_z)$ and the site $j = (j_x, j_y, j_z)$ when two resistors are removed.

First of all, let us consider the case when the resistor between $i_o$ and $j_o$ is missing. This case has been studied by many authors [2,7,8,13] where they express the equivalent resistance (i.e. $R_{o1}(i,j)$) in the perturbed lattice (i.e. the bond $(i_o, j_o)$ is removed) in terms of the perfect resistance (i.e. $R_o(i,j)$) as:

$$\frac{R_{o1}(i,j)}{R} = R_o(i,j) + \frac{[R_o(i,j_o) + R_o(j,i_o) - R_o(i,i_o) - R_o(j,j_o)]^2}{4[1 - R_o(i_o, j_o)]}. \tag{9}$$

For large separation between the sites $i$ and $j$, the above equation becomes [2,7,8,13]:

$$\frac{R_{o1}(l,m,n)}{R} \to \frac{R_o(i,j)}{R} = g_o = 0.505462. \tag{10}$$

Now, let us consider the case where the resistor between the sites $i_o$ and $j_o$ is removed in addition to the resistors between the sites $k_o$ and $l_o$. Here, we have to follow the same procedure presented above when one bond is only removed. The equivalent resistance (i.e. $R(i,j)$) between the two lattice sites $i = (i_x, i_y, i_z)$ and $j = (j_x, j_y, j_z)$ when the two bonds between the sites $(i_o, j_o)$ and $(k_o, l_o)$ are removed can be written as [13]:

$$R(i,j) = R_{o1}(i,j) + \frac{R}{1 - \frac{R'_{o1}(k_o,l_o)}{R}} \left\{ \frac{R_o(j,k_o) + R_o(i,l_o) - R_o(j,i_o) - R_o(i,k_o)}{2R} + \right.$$

$$\left. \frac{1}{1 - \frac{R_o(i_o,j_o)}{R}} \left( \frac{2R_o(i,j_o) - R_o(i,i_o) - R_o(j,j_o)}{2R} \right) \left\{ \frac{R_o(i_o,k_o) + R_o(i_o,l_o) - R_o(j_o,l_o) - R_o(j_o,k_o)}{2R} \right\} \right\}^2. \quad (11)$$

where $R_{o1}(i,j)$ is defined in Eq. (9).
and
$R'_{o1}(k_o,l_o)$ is the resistance between the ends of the removed bond $(k_o l_o)$ as affected from the removed bond $(i_o j_o)$, and from Eq. (9) one can write it as:

$$\frac{R'_{o1}(k_o,l_o)}{R} = R_o(k_o,l_o) + \frac{[R_o(k_o,j_o) + R_o(l_o,i_o) - R_o(k_o,i_o) - R_o(l_o,j_o)]^2}{4[1 - R_o(i_o,j_o)]}. \quad (12)$$

To check our result, take $k_o \to 0$ and $l_o \to 0$, then Eq. (11) reduces to Eq. (9). This means that the two broken bonds problem reduces to the problem of one broken bond.

It is important to study the asymptotic behavior of the resistance as the separation between $i$ and $j$ goes to infinity. In the case of the two removed resistors one can easily show that Eq. (11) goes to:

$$R(i,j) \to R_{o1}(i,j) \to R_o(i,j) = g_o = 0.505462. \quad (13)$$

**IV- Numerical Results and Discussion**

In this section, numerical results are presented for an infinite SC lattice including the perfect and both of the perturbed cases. The resistance between the sites $i = (0,0,0)$ and $j = (j_x, j_y, j_z)$ in an infinite perfect SC lattice is calculated in Asad et. al[8] (i.e. where various values for $\rho_1$, $\rho_2$ and $\rho_3$ are presented in Table 1 in Ref[8]).

For the case of one broken bond, one has to specify exactly the two ends of the removed bond and then the values of the perturbed resistance can be calculated using the calculated values of the perfect SC lattice (i.e. $R_o(i,j)$) and Eq. (9). Asad et al[8] study this problem where they consider two cases: First, the bond between $i_o = (0,0,0)$ and $j_o = (1,0,0)$ is removed. Second, then the removed bond is shifted and set between $i_o = (1,0,0)$ and $j_o = (2,0,0)$. They calculated the resistance between the sites $i = (0,0,0)$ and $j = (j_x, j_y, j_z)$ along the directions [100], [010] and [111]. Their calculated (i.e. $R_{o1}(i,j)$) values are arranged in Table 2 and Table 3 (i.e. see Ref[8].).

In the extended perturbed case where two bonds are broken, one has to specify the ends of the removed bonds $(i_o j_o)$ and $(k_o l_o)$ then using Eq. (11) and the calculated values of $R_o(i,j)$ and $R_{o1}(i,j)$ to calculate the values of the resistance ($R(i,j)$) in the new perturbed SC lattice. In this work we considered three cases: first, when the first removed bond is between $i_o = (0,0,0)$ and $j_o = (1,0,0)$, whereas the second broken bond is between $k_o = (1,0,0)$ and $l_o = (2,0,0)$. Our calculated values for the perturbed resistance (i.e. $R_1(i,j)$) are arranged in Table I below. In the second case, the first removed bond is between $i_o = (0,0,0)$ and $j_o = (1,0,0)$, whereas the second broken bond is between $k_o = (2,0,0)$ and $l_o = (3,0,0)$. Our calculated values for the perturbed resistance (i.e. $R_2(i,j)$) are arranged in Table I below. Finally, we consider the case where the first removed bond is between $i_o = (1,0,0)$ and $j_o = (2,0,0)$, whereas the second broken bond is between $k_o = (2,0,0)$ and $l_o = (3,0,0)$. Again, our calculated values for the perturbed resistance (i.e. $R_3(i,j)$) are arranged in Table I below.

In Figs. 1-3, the resistance for the perfect and the above three perturbed cases are plotted as a function of $j_x$. One can see that the equivalent resistance when two resistors are removed is always larger than that when only one resistor is removed. This is due to the positivity of the second term in Eq. (7). This also mean that the required resistance in the case where two resistors are broken is always larger than the resistance in the perfect lattice, and in general, one can say that; as the number of removed resistors increases in an infinite SC lattice the perturbed resistance increases.

Finally, from Figs. 1-3 one can see that the resistance in a perturbed infinite SC lattice is not symmetric under the transformation $j_x \to -j_x$. This is due to the fact that the inversion symmetry of the infinite lattice has been broken. Also, as the separation between the sites $i = (0,0,0)$ and $j = (j_x, 0, 0)$ increases then the equivalent resistance of the perturbed lattice tends to that of the perfect lattice.

**Table Captions**

**Table. 1** Calculated values for the resistance of an infinite SC lattice between the sites $i = (0,0,0)$ and $j = (j_x, 0, 0)$, for a perfect lattice ($R_o(i,j)/R$); Perturbed lattice due to removing the bonds resistors between (0,0),(1,0) and (1,0), (2,0)- ($R_1(i,j)/R$) -; Perturbed lattice due to removing the bonds resistors between (0,0),(1,0) and (2,0), (3,0)- ($R_2(i,j)/R$) - and Finally, perturbed lattice due to removing the bonds resistors between (1,0),(2,0) and (2,0), (3,0)- ($R_1(i,j)/R$) -.

Table. 1

| $j = (j_x, j_y, j_z)$ | $R_1(i,j)/R$ | $R_2(i,j)/R$ | $R_3(i,j)/R$ | $R_o(i,j)/R$ |
|---|---|---|---|---|
| (1,0,0) | 0.500519 | 0.500003 | 0.501203 | 0.333333 |
| (2,0,0) | 0.649239 | 0.513809 | 0.489077 | 0.419683 |
| (3,0,0) | 0.523619 | 0.555946 | 0.585457 | 0.450371 |
| (4,0,0) | 0.523162 | 0.517208 | 0.52499 | 0.464885 |
| (5,0,0) | 0.527225 | 0.5194 | 0.523558 | 0.473263 |
| (6,0,0) | 0.530916 | 0.52325 | 0.526388 | 0.478749 |
| (7,0,0) | 0.533926 | 0.526501 | 0.52922 | 0.482685 |
| (0,0,0) | 0 | 0 | 0 | 0 |
| (-1,0,0) | 0.358484 | 0.356764 | 0.358965 | 0.333333 |
| (-2,0,0) | 0.455112 | 0.454693 | 0.456891 | 0.419683 |
| (-3,0,0) | 0.48983 | 0.489203 | 0.491365 | 0.450371 |
| (-4,0,0) | 0.505751 | 0.505226 | 0.507348 | 0.464885 |
| (-5,0,0) | 0.521301 | 0.514295 | 0.516399 | 0.473263 |
| (-6,0,0) | 0.52706 | 0.520128 | 0.522153 | 0.478749 |
| (-7,0,0) | 0.530963 | 0.524033 | 0.525774 | 0.482685 |

# Figure Captions

**Fig. 1** The resistance between $i=(0,0,0)$ and $j=(j_x,0,0)$ along [100] direction of the perfect (circles) and the perturbed infinite SC lattice (squares) as a function of $j_x$. The ends of the removed resistors are $i_o=(0,0,0)$ and $j_o=(1,0,0)$, $k_o=(1,0,0)$ and $l_o=(2,0,0)$.

**Fig. 2** The resistance between $i=(0,0,0)$ and $j=(j_x,0,0)$ along [100] direction of the perfect (circles) and the perturbed infinite SC lattice (squares) as a function of $j_x$. The ends of the removed resistors are $i_o=(0,0,0)$ and $j_o=(1,0,0)$, $k_o=(2,0,0)$ and $l_o=(3,0,0)$.

**Fig. 3** The resistance between $i=(0,0,0)$ and $j=(j_x,0,0)$ along [100] direction of the perfect (circles) and the perturbed infinite SC lattice (squares) as a function of $j_x$. The ends of the removed resistors are $i_o=(1,0,0)$ and $j_o=(2,0,0)$, $k_o=(2,0,0)$ and $l_o=(3,0,0)$.

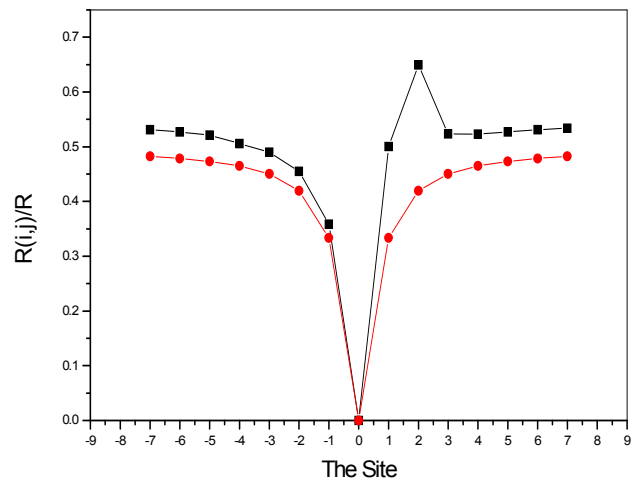

Fig. 1

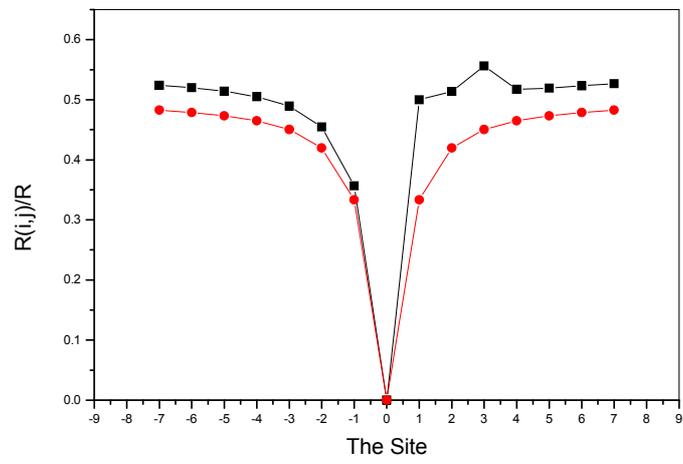

Fig. 2

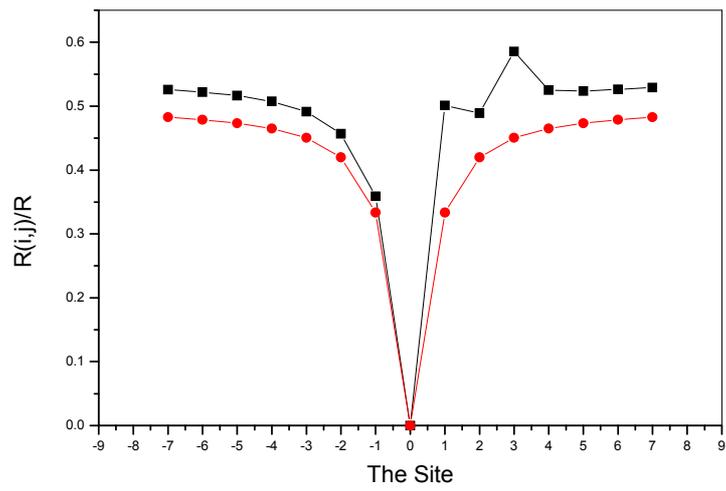

Fig. 3